\documentclass{acmart}
\usepackage{hyperref}
\usepackage{hyperxmp}

\AtBeginDocument{%
  \providecommand\BibTeX{{%
    \normalfont B\kern-0.5em{\scshape i\kern-0.25em b}\kern-0.8em\TeX}}}

\usepackage{url}
\usepackage{xurl}

\usepackage{ragged2e}
\usepackage{etoolbox} 

\usepackage{subcaption}
\usepackage{mdframed}

\usepackage{siunitx,etoolbox}
\usepackage{dcolumn,booktabs}
\usepackage{multirow}

\usepackage{array}
\newcolumntype{C}[1]{>{\centering\arraybackslash}p{#1}}
\usepackage{bm}
\usepackage{listings}
\usepackage{xltabular}
\usepackage{tabularray}

\usepackage{amssymb} 
\usepackage{pifont}  

\usepackage{tikz}
\usepackage{xcolor}

\newmdenv[
  leftmargin=2cm,
  rightmargin=2cm,
  backgroundcolor=gray!20,
  linewidth=1pt,
  linecolor=black,
  innertopmargin=10pt,
  innerbottommargin=10pt,
  innerleftmargin=15pt,
  innerrightmargin=15pt
]{indentedbox}

\definecolor{oxfordblue}{rgb}{0.0, 0.13, 0.28}
\definecolor{harvardcrimson}{rgb}{0.79, 0.0, 0.09}
\definecolor{dartmouthgreen}{rgb}{0.05, 0.5, 0.06}
\definecolor{princetonorange}{rgb}{1.0, 0.56, 0.0}
\definecolor{yaleblue}{rgb}{0.06, 0.3, 0.57}
\definecolor{usccardinal}{rgb}{0.6, 0.0, 0.0}
\definecolor{uclablue}{rgb}{0.33, 0.41, 0.58}
\definecolor{msugreen}{rgb}{0.09, 0.27, 0.23}
\definecolor{cornellred}{rgb}{0.7, 0.11, 0.11}
\definecolor{pomegranate}{RGB}{192, 57, 43}
\definecolor{anti-pomegranate}{RGB}{43,178,192}
\definecolor{alizarin}{RGB}{231, 76, 60}
\definecolor{anti-belize}{RGB}{185, 41, 56}
\definecolor{belize}{RGB}{41, 128, 185}
\definecolor{peter}{RGB}{52, 152, 219}
\definecolor{green}{RGB}{22, 160, 133}
\definecolor{anti-green}{RGB}{160,22,118}
\definecolor{turquoise}{RGB}{26, 188, 156}
\definecolor{pumpkin}{RGB}{211, 84, 0}
\definecolor{anti-pumpkin}{RGB}{0,22,211}
\definecolor{carrot}{RGB}{230, 126, 34}
\definecolor{wisteria}{RGB}{142, 68, 173}
\definecolor{anti-wisteria}{RGB}{99,173,68}
\definecolor{amethyst}{RGB}{155, 89, 182}
\definecolor{nephritis}{RGB}{39, 174, 96}
\definecolor{anti-nephritis}{RGB}{174,39,117}
\definecolor{grey-bg}{RGB}{242,242,242}

\newcolumntype{d}[1]{D{.}{.}{#1}}
\newrobustcmd*{\bftabnum}{%
	\bfseries
	\sisetup{output-decimal-marker={\textmd{.}}}%
}

\makeatletter

\begin{document}

\title[Generative AI Literacy]{Generative AI Literacy: A Comprehensive Framework for Literacy and Responsible Use}

\author{Chengzhi Zhang}
\email{czhang694@gatech.edu}
\orcid{0000-0002-6868-2285}
\affiliation{
  \institution{Georgia Institute of Technology}
  \city{Atlanta}
  \state{Georgia}
  \country{USA}
}

\author{Brian Magerko}
\email{magerko@gatech.edu}
\orcid{0000-0003-1900-4020}
\affiliation{
  \institution{Georgia Institute of Technology}
  \city{Atlanta}
  \state{Georgia}
  \country{USA}
}

%
\renewcommand{\shortauthors}{Zhang \& Magerko}

\begin{abstract}

After the release of several widely adopted artificial intelligence (AI) literacy guidelines by 2021, the unprecedented rise of generative AI since 2023 has transformed the way we work and acquire information worldwide. Unlike traditional AI algorithms, generative AI exhibits distinct and more nuanced characteristics. However, a lack of robust understanding of generative AI hinders individuals' ability to use generative AI effectively, critically, and responsibly, which we can call \textit{generative AI literacy}. To address this gap, we reviewed and synthesized existing literature and proposed \textit{generative AI literacy} guidelines with 12 items organized into four aspects: (1) generative AI tool selection and prompting, (2) understanding interaction with generative AI, (3) understanding generative AI outputs, and (4) high-level understanding of generative AI technologies. These guidelines aim to support schools, companies, and organizations in developing frameworks that support their members to use generative AI in an efficient, ethical, and informed way. 

\end{abstract}




\maketitle
\section{Introduction}

Generative AI---broadly referring to large-scale, pretrained generative models, such as foundational and large language models (LLMs)---has gained unparalleled worldwide popularity over the past several years. This interest surged into a global phenomenon following the launch of ChatGPT-3.5 in November 2022~\cite{Mollman_2022}. After the release of the ``Five Big Ideas in AI'' from the AI4K12 organization in 2019~\cite{touretzky2019envisioning}, followed by AI literacy competency frameworks proposed by Long \& Magerko in 2020~\cite{long2020ai} and Ng et al. in 2021~\cite{ng2021conceptualizing}, the field of artificial intelligence (AI) has undergone a significant transformation with the rise of generative AI~\footnote{We used the term ``generative AI'' for accessibility and to align with common discourse~\cite{mittal2024comprehensive, annapureddy2024generative}. While more precise terms include ``Large-Pretrained Generative Models (LPGMs),'' this chosen term retains technical relevance for a broader learner audience.}, especially since the worldwide boom of ChatGPT in early 2023~\footnote{ChatGPT: \url{https://chatgpt.com/}}, followed by other large language generative AI applications like Copilot~\footnote{Copilot: \url{https://copilot.microsoft.com/}}, Gemini~\footnote{Gemini: ~\url{https://gemini.google.com/app}} and Deepseek~\footnote{Deepseek: \url{https://www.deepseek.com/}}. Unlike traditional AI algorithms like decision trees, semantic networks, and convolutional neural networks (CNNs)---which were captured in existing AI literacy guidelines chartered before this boom---generative AI exhibits distinct and specialized characteristics that are not quite addressed by popular guidelines. For instance, generative AI has a limited context window, can output falsified facts (often called \textit{hallucinations}) in scale, and causes significant environmental impact from model training and inferencing. These characteristics are inherent in generative AI, and are not mentioned in current high-level AI literacy guidelines~\cite{long2020ai,ng2021ai}. Furthermore, generative AI has demonstrated advanced multilingual comprehension capabilities~\cite{ahuja2023mega} and human-level performance on many cognitive aptitude tests---including General Language Understanding Evaluation (GLUE) and Stanford Question Answering Dataset (SQuAD) test~\cite{ray2023chatgpt}. These demonstrated human/superhuman capabilities often lead users to overestimate the intelligence of generative AI systems, a phenomenon known as the ``Eliza effect''~\cite{hofstadter1995ineradicable}---tendency to attribute greater intelligence to responsive computer programs than they actually possess~\cite{switzky2020eliza}. Such overconfidence in generative AI's abilities has resulted in prevalent misconceptions about its limitations~\cite{johri2024misconceptions,kidd2023ai}. These misunderstandings can have far-reaching consequences across various domains. For instance, they can erode professionalism in the workplace~\cite{NEUMEISTER_2023_lawyers}, compromise integrity in academic settings, including scientific writing~\cite{alkaissi2023artificial}, and even result in fatal outcomes for both adults~\cite{xiang_2023_he} and minors~\cite{Atillah_2023_AI, Duffy_2025}.

The unique characteristics of generative AI---differentiating from general AI models---and the potential harms it can cause due to public misconceptions underscore the need for specialized literacy guidelines tailored to generative AI. Unlike AI algorithms like decision trees that discriminate and classify (e.g. spam filters) and convolutional neural networks (CNNs) that produce deterministic results (e.g. generating text from speech), mainstream large-pretrained generative models often rely on advanced transformer architectures~\cite{vaswani2017attention}, which enabled exceptional performance and also introduced a high degree of non-determinism and variability in their outputs. However, the existing AI literacy guidelines often fail to address these specificities, as they primarily focus on non-generative AI systems.

Current literature on ``generative AI literacy'' is limited but growing. Examples include defining twelve competencies for generative AI literacy~\cite{annapureddy2024generative}, outlining four generative AI literacy dimensions including ``knowledge,'' ``application,'' ``evaluation,'' and ``ethics''~\cite{o2024factors}, drafting the Generative AI Literacy Assessment Test (GLAT)~\cite{jin2024glat}, and hosting workshops for educating students on using generative AI~\cite{sullivan2024improving}. While the outlined twelve competencies provide a high-level framework~\cite{annapureddy2024generative} (e.g. ``3. Knowledge of the capabilities and limitations of generative AI tools''), we further consolidated key points from existing literature into more concrete guidelines for policymakers and educators to design effective learning interventions. Our guidelines are comprised of four key aspects of interacting with generative AI: (1) generative AI tool selection and prompting, (2) understanding interaction with generative AI, (3) understanding generative AI outputs, and (4) high-level understanding of generative AI technologies. This work offers practical guidelines for using generative AI technology in an effective, ethical, and informed manner in order to bridge the gap between rapid advances in generative AI and the public's proficiency.

\section{Defining Generative AI Literacy}

\paragraph{What is generative AI?}

\textit{Generative AI} refers to AI models that are capable of generating new, meaningful content such as text, images, code, audio, music---from typically large amounts of training data~\cite{feuerriegel2024generative, muller2024genaichi}. Foster's book provides a more comprehensive visualization of the history of generative AI, categorizing its development and highlighting representative products up to early 2023~\cite{foster2023generative}. Since the emergence of AI as a field in the 1950s~\cite{haenlein2019brief}, generative AI like VAE and GAN models gained prominence in the 2010s, excelling in image generation tasks. However, the introduction of the transformer architecture in 2017~\cite{vaswani2017attention} marked a new turning point, leading to the development of LLMs like ChatGPT, and sparking a global surge in generative AI applications.

We conducted an exploratory review of existing literature to inform our generative AI literacy guidelines. Given the rapidly evolving nature of generative AI, we anticipate that future researchers will subsequently expand upon this work to develop more refined and up-to-date guidelines (e.g. generative AI could soon be embodied or embedded in classrooms). This framework serves as an initial step toward developing more comprehensive resources for educators, policymakers, and researchers to design learning interventions---such as curricula, workshops, and other tools---aimed at improving the public's understanding of generative AI. We define generative AI literacy as:

\begin{quote}
    A set of guidelines for designing learning interventions aimed at enhancing users' understanding of generative AI, enabling them to interact with it effectively, responsibly, and critically. These guidelines are intended to assist educators, schools, companies, and organizations in developing frameworks that empower their members, such as students, employees, and stakeholders, to use generative AI in ethical and informed ways.
\end{quote}

The most widely used generative AI is the generative language model---commonly referred to as large language models (LLMs) or large, pre-trained language models (PLMs)~\cite{min2023recent}. While generative AI broadly includes applications for video generation~\cite{li2018video}, image generation~\cite{liu2022opal}, design generation~\cite{yan2022toward}, and code generation~\cite{sun2022investigating}, as detailed above, our guidelines primarily focus on text-based generative models. These models are the most widely used and studied, though many of the guidelines can also apply to other types of generative AI systems.
\section{Method}

While systematic literature reviews are commonly used to identify practices in well-established research fields, generative AI literacy is a nascent research area with limited coverage in existing literature. To address this gap, we employed a scoping study to systematically gather and develop a comprehensive set of guidelines. The scoping study is \textit{``an approach to reviewing the literature which to date has received little attention in the research methods literature''}~\cite{arksey2005scoping}, and it has previously been adopted in defining AI literacy competencies~\cite{long2020ai}. Differentiating from a systematic review aiming at ``searching for particular study designs'', the goal of scoping studies is to have ``in‐depth and broad results''~\cite{arksey2005scoping}, ensuring a comprehensive coverage of existing literature. 

In this active searching process, guided by our definition of generative AI literacy, we formulated two key inclusion criteria to direct our search: (1) Is this knowledge essential for generative AI users with a non-technical background? (2) Will including this in the guideline help learners interact with generative AI more effectively, ethically, and responsibly? 

In the search process, we began by querying the ACM library, the Springer library, and Google Scholar using the keywords such as ``generative AI literacy'', ``generative AI characteristics'', and ``generative artificial intelligence literacy''. However, the preliminary search only yielded a limited set of literature. Thus, we expanded our search to include both peer-reviewed articles and grey literature (e.g. government reports, policy documents, and working papers) while incorporating additional keywords including ``generative AI'', ``ChatGPT'', ``LLM'', ``foundational models'', ``guideline'', and ``use wisely''. To enhance the comprehensiveness of our search, we incorporated two AI-empowered academic search tools--Semantic Scholar~\footnote{Semantic Scholar: \url{https://www.semanticscholar.org/}} and Elicit~\footnote{Elicit: \url{https://elicit.com/}}. These tools further enabled us to identify literature that was semantically relevant but did not necessarily include those keywords we queried. For example, the book \textit{Generative Artificial Intelligence: What Everyone Needs to Know}~\cite{kaplan2024generative}, which provides guidelines for generative AI users, was discovered through semantic search and did not appear in keyword-based queries.

Additionally, we incorporated snowball sampling, a technique recommended in scoping study processes~\cite{arksey2005scoping}---examining the bibliographies of relevant studies to identify additional sources~\cite{parker2019snowball}. This approach allowed us to gather a diverse range of materials, including peer-reviewed, generative AI use guidelines from accredited university libraries~\cite{harvard2023,ubc_genai2024,hku_ai2023,hkust2024generativeai}, international non-profit organizations like The United Nations Educational, Scientific and Cultural Organization (UNESCO)~\cite{unesco_chatgpt2023}, the Council of the European Union~\cite{counseleu2023chatgpt}, and a public-facing course from a globally recognized leader in AI~\cite{generative_ai_for_everyone}. This initial search and subsequent refinement resulted in a preliminary set of guidelines, with each item supported by at least one relevant literature.

Following the initial search, the first author (who has an educational background in AI) conducted an exploratory reading process, reviewing the abstracts and skimming the contents of the collected literature. For relevant sections, the first author conducted a thorough reading and labeled each literature according to the draft guideline items. Throughout this process, the first author consulted the second author (who has more than 25 years of experience in AI research) for professional input and advice. This involves adding, consolidating, or removing draft guideline items. Additionally, we searched for relevant literature to support and refine the guidelines until reaching a literature saturation. In total, we reviewed 100 literature and developed a set of 12 guidelines, with each literature mapped to at least one guideline item. The Table \ref{tab:lit_review} lists the categories of literature reviewed. For consistency and to emphasize the educational focus of the guidelines, we use the term ``learners'' throughout the paper to refer to users of generative AI who are in need of AI literacy skills. In the following discussion, we use the notation ``G1'', ``G2'', etc., to refer to each guideline item.

\begin{table*}
    \small
    \caption{Literature reviewed sorted by venue type}
    \Description{ }
    \label{tab:lit_review}
    \begin{tblr}{
      colspec={l|l},
      row{odd}={bg=grey-bg},  
      row{1}={bg=black,fg=white},
    }
        Venue Type &  \\
        Conference papers & 23 \\
        Journal papers & 28 \\
        Books & 4 \\
        Other grey literature & 45\\
    \end{tblr}
\end{table*}


\section{Guidelines for Generative AI Tool Selection and Prompting}

\begin{itemize}
    \item \textbf{G1: Learners must carefully determine whether they should use generative AI for their task and which tool to select.}
    \textit{(Supporting References:}~\cite{cox2024algorithmic,hkust2024generativeai,ray2023chatgpt, unesco_chatgpt2023})
\end{itemize}

Learners need to assess the appropriateness of using generative AI at the forefront, differentiating from high-risk contexts and low-stakes applications such as composing non-urgent emails. Specifically, the United Nations Educational, Scientific, and Cultural Organization (UNESCO) provides a flow chart to help determine the safe use of ChatGPT~\cite{unesco_chatgpt2023}: if users can not verify the accuracy of the output and correctness is critical, then it is unsafe to use ChatGPT. Universities have varying stances on using generative AI. Some are embracing it as an educational tool and offering courses like ``Facilitating Learning and Research with Generative AI''~\cite{hku_ai2023} while others consider submitting AI-generated content in whole or in part as cheating~\cite{gwu_ai_guidelines}. In educational and workplace settings, learners should consult the institution's policy and evaluate the context before using generative AI. Beyond simply adhering to policy, learners should understand while generative AI tools (e.g. Elicit) can help identify learning resources such as relevant literature, they can not replace human thinking and learning processes~\cite{Chow_2025}. So in both educational and workplace settings, learners should employ these tools with skepticism. 

For non-professional or creative purposes like image and music generation, learners should compare the capabilities of various tools to select the most suitable one. Different generative AI applications have distinct capacities, and reference sheets listing key features---such as advantages, disadvantages, privacy policy, and trained data of key generative AI applications---could help learners determine the most suitable tool~\cite{hkust2024generativeai,ray2023chatgpt}.  In summary, learners should consider their goals, the institutional policies, use contexts, and tool performance to decide whether they should use generative AI and which generative AI tool to use.

\begin{itemize}
    \item \textbf{G2: Learners need to develop safe and effective prompting skills, ensuring user privacy while interacting effectively.}
    \textit{(Supporting References:}~\cite{unesco_chatgpt2023,harvard2023, ubc_genai2024,texas_tech2023, hku_ai2023, jin2025generative, annapureddy2024generative, sullivan2024improving, generative_ai_for_everyone})
\end{itemize}

\paragraph{How do people prompt generative AI?}

Existing research with LLM users revealed their prompting characteristics: (1) they were opportunistic but not systematic, (2) they tended to over-generalize the responses, (3) they held the inappropriate mental model towards LLMs, derived from human-to-human instructional experience~\cite{zamfirescu2023johnny}. For generative image AI, research also witnessed users' inappropriate mental models---including expecting generative image AI to interpret prompts like humans do---leading to undesired outputs~\cite{mahdavi2024ai, zhang2023generative}. Apart from these two studies with adults, a study with children also witnessed that children interact with generative AI chatbots as if they were interacting with people (e.g. inputting ``Are you a boy or girl?'')~\cite{belghith2024testing}. These all revealed the learners' misconceptions about and ineffective prompting techniques for generative AI, highlighting the need for adequate prompting skills.

The premise for effective prompting involves inputting generative AI---like chatbots---with questions they are capable of answering. A well-accepted framework is the CLEAR prompting guideline~\cite{lo2023clear}, which stands for ``Concise'', ``Logical'', ``Explicit'', ``Adaptive'' and ``Reflective'', and it has been endorsed by university libraries~\cite{harvard2023, texas_tech2023}. The guideline emphasizes not only crafting a single well-structured prompt but also iteratively refining based on the outputs. In the context of generative image AI, researchers have developed specialized guidelines for effective image generation~\cite{liu2022design}. Some products (e.g. Dall·E) have further released their product-specific guidelines~\cite{Dalle2PromptBook}. Prompting generative image AI requires additional expertise, including adjusting the parameters and seeds to achieve the desired results~\cite{xu2024seed}. Beyond effectiveness, user privacy and security are critical considerations. To safeguard data privacy, institutions advise against inputting personal data~\cite{texas_tech2023} and recommend privacy-protected AI tools~\cite{wellington2023}.
\section{Guidelines for Understanding Interaction with Generative AI}

\begin{itemize}
    \item \textbf{G3: Generative AI has a short-term memory (i.e. the context window), so users need to re-frame or repeat context as needed.}
    \textit{(Supporting References:}~\cite{zamfirescu2023johnny, jin2024glat, noever2023llm, ibrahimzada2024program})
\end{itemize}

The limited context window (in more accessible words, short-term memory) is a significant challenge for generative AI, especially LLMs~\cite{ibrahimzada2024program}. This limitation causes generative AI to forget earlier inputs during long conversations and ``struggle to maintain continuity within long dialogs''~\cite{noever2023llm}. This constraint was also highlighted in the Generative AI Literacy Assessment Test (GLAT)~\cite{jin2024glat}. Although natural language processing (NLP) researchers are actively working to extend these context windows~\cite{ding2024longrope,ibrahimzada2024program}, this issue persists.  Empirical studies reveal that users who are unaware of this constraint often experience frustration~\cite{zamfirescu2023johnny}. To avoid frustration, users must recognize that, unlike conversing with humans, generative AI does not retain context automatically; continuity must be maintained by manually re-framing and repeating information periodically.

\begin{itemize}
    \item \textbf{G4: Generative AI lacks social cognition but can demonstrate fake empathy and theory of mind. Learners should avoid overrelying on it as a social companion.}
    \textit{(Supporting References:}~\cite{cuadra2024illusion,zhou2023far,noever2023llm, saritacs2025systematic, wang2021towards})\looseness=-1
\end{itemize}

As the most widely adopted form of generative AI, LLMs generate word sequences with probabilistic approaches. While LLM applications can simulate human-like responses~\cite{park2023generative}, they lack the capacity for social cognition, including theory of mind (ToM)~\cite{zhou2023far, deshpande2024embracing}. As defined by Premack, ToM refers to the ability to attribute mental states---such as beliefs, intentions, and emotions---to oneself and others, and to use that understanding for behavior predictions~\cite{premack1978does}. While Google's Language Model for Dialogue Applications (LaMDA) reportedly passed a famous ToM assessment test~\cite{noever2023llm}---which is a hallmark of consciousness---the result remains controversial. Researchers employed techniques to make LLMs appear as if they possess ToM~\cite{zhou2023far}, but this remains a superficial imitation rather than a genuine understanding (i.e. ``LLMs manipulate vectors that do not seem grounded in the world and thus lack intrinsic meaning''~\cite{mollo2023vector}). An up-to-date pre-print analyzing existing literature concluded that although new LLMs demonstrated advanced ToM abilities, they still rely on spurious correlations instead of a solid understanding~\cite{saritacs2025systematic}, further challenging LLMs' ToM abilities. 

LLMs also could not engage in \textit{shared mental state construction} (SSM), a concept described in the \textit{International Encyclopedia of the Social \& Behavioral Sciences} as the ability of team members to use shared knowledge to predict task needs and anticipate others' actions, enabling adaptive behavior~\cite{smelser2001international}. Generative AI lacks the capacity for participatory sense-making---dynamic meaning generation and transformation through interaction~\cite{de2007participatory, davis2016empirically}---showing a more general lack of understanding of social cognitive processes. Despite these shortcomings, people often turn to AI for on-demand social validation~\cite{turkle2024we}, which can reinforce people's existing biases or even trap users in the ``bias bubbles''. The demonstrated fake empathy would create an illusion of connection and understanding~\cite{cuadra2024illusion}, and possibly increase the vulnerability of people interacting with LLMs. So far, it has led teenagers into emotionally and sexually abusive relationships, and caused fatal consequences for both minors~\cite{cnn2024chatbot, Duffy_2025} and adults~\cite{brusselstimes2023}. These limitations highlight LLMs' inability to engage in genuine social cognitive processes and caution users against over relying on them as social companions.

\section{Guidelines for Understanding Generative AI's Output}

\begin{itemize}
    \item \textbf{G5: Generative AI can output content and instructions that are harmful to people, so learners should treat the outputs with caution and skepticism.}
     \textit{(Supporting References:}~\cite{unesco_chatgpt2023, ubc_genai2024, fabricating2023, chatbot_lawsuit2024, wellington2023})
\end{itemize}

Even as generative AI has demonstrated state-of-the-art superhuman fact-retrieval abilities, learners need to understand that generative AI outputs are not always benign. LLMs deploy probabilistic models, and they do not know the real-world implications of output contents. It can produce harmful, inappropriate, deceptive, and misleading content. For instance, there are reports of teenagers and adults using generative AI chatbots outputting dangerous instructions, such as those leading to electric shocks~\cite{bbc2021alexa} or misleading advice with severe ethical implications, like instructions to harm their guardians~\cite{chatbot_lawsuit2024}. To mitigate these risks, some companies have adopted red-teaming practices before their generative AI models' launch, and shared related tools with developers building generative AI applications on their company platform~\cite{microsoft2024transparency, bundschuh2023pyrit}. While these measures help assess and reduce the likelihood of producing harmful content, they cannot entirely eliminate the risks. Learners must remain vigilant about the potential for harmful outputs.

\begin{itemize}
    \item \textbf{G6: Generative AI can output falsified facts (often called \textit{hallucinations)}, so learners should critically consume the results and check for correctness, especially in high-stakes contexts.} 
    \textit{(Supporting References:}~\cite{park2024ai, walters2023fabrication, nyt_disinformation2024, uci_2023, rochester2023, northwestern_2024, jin2024glat})
\end{itemize}

Differentiating from harmful contents---that emphasize non-false but dangerous contents that can directly harm individuals or groups---``falsified facts''~\footnote{In some literature, people term this the hallucination effect~\cite{del2022machine}. However, according to the \textit{American Psychology Association (APA)}, ``hallucination'' means ``a false sensory perception that has a compelling sense of reality despite the \textbf{absence} of an external stimulus.'' This deviates from the ``seemingly true'' but false information, ``factually incorrect'' and ``fabricated information'' that learners encounter with AI-generated content. } focus on ``plausible falsehoods''~\cite{zhang2025navigating}, including misinformation. These are the contents that are seemingly true (e.g. fabricated scholarly references), but verifiable if learners validate their accuracy. This aligns with longstanding calls in academic literature for critical information literacy~\cite{brisola2019critical} and critical media literacy~\cite{kellner2005media}, which emphasize equipping users with the skills to interrogate the authenticity of content. The \textit{American Psychology Association (APA)} defines misinformation as ``false or inaccurate information—getting the facts wrong'', and disinformation as ``false information which is deliberately intended to mislead--intentionally misstating the facts''~\cite{misinformation2023}. As a widely used generative AI tool, ChatGPT has gained notoriety for generating fabricated academic citations~\cite{alkaissi2023artificial}, misleading narratives~\cite{hsu_2023_disinformation}, and misinformation like bogus case law~\cite{NEUMEISTER_2023_lawyers}. These all pose significant challenges to people who use ChatGPT as a ``search engine'' without knowing those true limitations. Learners should distinguish between low-stakes scenarios (e.g. students using generative AI for inspiration or drafting informal emails) and high-stakes scenarios (e.g. lawyers use generative AI for finding specific legal clauses). In high-stakes situations, regular fact-checking is essential to mitigate risks. While AI-generated misinformation is not a new issue, the ease, speed, and credibility with which generative AI can produce such content are unprecedented~\cite{zhou2023synthetic}. Due to their speed, researchers are concerned that LLMs might pollute our information ecosystems~\cite{sobieszek2022playing, manovich2018ai}, exacerbating the spread of misinformation. 

\begin{itemize}
    \item \textbf{G7: Generative AI can lack explainability of its results, so learners should validate its sources when possible and always cross-check the results.}
    \textit{(Supporting References:}~\cite{ehsan_expanding_2021,jin2024glat,wineburg2017lateral,schneider2024explainable})
\end{itemize}

The lack of explainability in LLMs---rooted in the complex mathematical nature of deep neural networks---renders their decision-making processes a ``black box'' to learners. Consequently, learners must fact-check important information generated by these systems. While tools like Deepseek (with web-search feature), as well as academic tools such as Sourcely~\footnote{Sourcely: \url{https://www.sourcely.net/}} and Elicit~\footnote{Elicit: \url{https://elicit.com/}}, can provide information sources, many generative models remain opaque and lack transparency in their outputs. 

Explainable AI system aims to address this issue through detailed explanations~\cite{gunning2019xai}. Explainable AI is critical for verifying information and providing traceable, investigable sources, with approaches that include technically visualizing neural networks---the foundational building blocks of generative AI~\cite{yan2021visualizing}, and constructing socially situated XAI systems~\cite{ehsan_expanding_2021} with automated explanations~\cite{ehsan2019automated}. However, these methods still fall short of the true ``explainability'' found in human-human interactions, pushing the responsibility on users to fact-check AI-generated results. \textbf{To ensure best accuracy, learners must compare AI outputs with ground truth information.} Some universities recommend strategies like ``lateral reading''~\cite{wineburg2017lateral}---cross-referencing multiple sources to verify information validity~\cite{slcc_2023}. This approach can help mitigate the risks associated with the opaqueness of generative AI systems. Furthermore, even if some systems can provide evidence, reasoning, and explanations---often presented as ``deep thinking'' and ``deep reasoning'' features---they can be an illusion~\cite{shojaee2025illusion}. Thus, learners need to scrutinize the coherence and validity of the reasoning presented. 

\begin{itemize}
    \item \textbf{G8: Generative AI can output biased results, so learners need to critically evaluate output for unmitigated bias.}
    \textit{(Supporting References:}~\cite{bender2021dangers, omiye2023large,bozkurt2024generative,gwu_ai_guidelines,ubc_genai2024,lo2023clear, machine_bias, carnegie_2025, coeckelbergh2020ai})
\end{itemize}

Bender et al. provide a detailed analysis of the origins of bias in AI systems from a computational linguistics perspective~\cite{bender2021dangers}. They highlight that bias stems from the datasets used to train these models, which often include biased information from online sources and lack media coverage for less-digitalized communities, etc. As a result, ChatGPT and similar applications can exhibit gender and racial bias~\cite{ovalle2023decoding}, cultural and linguistic bias~\cite{ray2023chatgpt}, and bias against individuals with disabilities~\cite{glazko2024identifying}, etc. So that these generative AI applications could not provide fair and balanced outputs. While bias in text generation can often be implicit, bias with text-to-image generators is more explicit. For example, prompting an image AI with ``Black African doctors providing care for white suffering children'' has produced images on which children are invariably black while doctors are white~\cite{drahl_2023_AI}. Learners should be aware of the bias in outputs and be critical of the generated results.

\section{Guidelines for High-level Understanding of Generative AI}
\begin{itemize}
    \item \textbf{G9: Digital and non-digital content encountered in daily lives can be AI-generated. Learners need to critically evaluate how genuine media content is.}
    \textit{(Supporting References:}~\cite{mollo2023vector,counseleu2023chatgpt,hailtik2023criminal,zhou2023synthetic,chadha2021deepfake})
\end{itemize}

The existing generative AI literacy guidelines~\cite{bozkurt2024generative} emphasize the ability to distinguish AI-generated content as a key competency. However, some argue that humans struggle to reliably distinguish AI-generated content~\cite{mla2024aiguide, chadha2021deepfake}, a challenge that is being exacerbated by the rapid advancement of generative AI. Therefore, we propose that users should focus more on \textbf{being aware of the existence of AI-generated content and being critical of what they encounter} rather than striving to distinguish all AI-generated content. This is particularly relevant in higher education, where AI-generated content detection tools are used despite their often being unreliable and prone to bias. Consequently, many higher institutions discourage the faculty and staff from using those detectors~\cite{ubc_genai2024,texas_tech2023}. Apart from AI-generated digital and non-digital contents~\cite{Salkowitz_2025}, the rise of generative AI has also heightened the risks of fraud, scams, and impersonation crimes through techniques like deepfaking~\cite{hailtik2023criminal,chadha2021deepfake}. These techniques make it easier for malicious actors to deceive individuals and organizations, underscoring the need for vigilance regarding the source of the content learners encounter.
 
While some organizations, including academic institutions~\cite{ubc_genai2024} and companies~\cite{microsoft2024transparency}, have transparently disclosed their use of generative AI, such responsible AI practices are not yet widespread. In everyday scenarios, such as customer service, the role of generative AI often goes unacknowledged. It highlights the importance of learners remaining cautious and critical of the genuineness of content encountered in the wild, underscoring the importance of fostering information and media literacy highlighted in G6.

\begin{itemize}
    \item \textbf{G10: Learners need to understand how generative models ``know'' (and ``don't know'') concepts compared to human knowing.} 
    \textit{(Supporting References:}~\cite{mollo2023vector, noever2023llm,counseleu2023chatgpt, mahowald2024dissociating})
\end{itemize}

Generative AI models are capable of ``knowing'' concepts and information, but not necessarily in the same way and depth as humans do. LLMs' being able to predict the word sequence does not necessarily mean they can engage in thinking and reasoning~\cite{mahowald2024dissociating, west2023generative}. Knowledge can be grounded in a cognitive system in referential, sensorimotor, relational, communicative, and epistemic ways~\cite{mollo2023vector}. But generative AI models---like LLMs---are typically trained with a humongous dataset from the internet, thus only grounded in linguistic and symbolic ways. The phenomenon has been captured in the famous ``Chinese room argument''~\cite{cole2004chinese}, which presents that they ``do not have the ability to process and understand the meaning in the way humans do''~\cite{counseleu2023chatgpt}. Generative image AI operates on visual data formats, but none of the existing generative AI to date has sensorimotor grounding in tactiles and smells (e.g. LLMs can provide step-by-step instructions on how to play violin, but they do not understand the true meaning of ``press the string against the fingerboard''). Understanding how a generative AI model ``knows'' things can be a critical skill in influencing how learners choose, interact with generative AI, and critically consume generative AI outputs.  

\begin{itemize}
    \item \textbf{G11: Learners need to be aware of the broader social impacts of generative AI, including hidden human labor, environmental impact, etc.} 
    \textit{(Supporting References:}~\cite{crawford2021atlas, lee2025impact, heikkila2022artist,unesco_chatgpt2023, jin2025generative, barrett2023not, dang2024ethical,generative_ai_for_everyone,weidinger2022taxonomy})
\end{itemize}

Beyond significant environmental costs of the computational power required both in model training and inferencing~\cite{berthelot2025understanding}, generative AI models also demand an immense amount of human labor. Crawford’s \textit{Atlas of AI} reveals that a vast support system is required to maintain and update these systems, like Amazon Echo~\cite{crawford2021atlas}, highlighting that their operation is not magical but deeply reliant on human effort. Moreover, the broader societal implications of generative AI are profound, including undermining critical thinking~\cite{lee2025impact,kasneci2023chatgpt}, infringing copyright~\cite{heikkila2022artist, deckker2025dreams}, reducing aesthetic diversity~\cite{manovich2018ai}, holding a fake promise of creativity~\cite{chakrabarty2024art}, homogenizing collective creativity~\cite{anderson2024homogenization, doshi2024generative}, and exacerbating educational inequalities~\cite{jin2025generative}, etc. Generative AI is also reshaping the skills needed in the workplace~\cite{mckinsey_ai_2025, epstein2023art}, underscoring the importance of learners' understanding of the wider generative AI social impacts. Learners should be aware of these consequences as they engage with generative AI.

\begin{itemize}
    \item \textbf{G12: The capabilities of generative AI evolve rapidly. Learners need to keep updated with its up-to-date capabilities and limitations.} 
    \textit{(Supporting References:}~\cite{openai2024chatgptsearch,google2024bardsearch,niszczota2023gpt,aiindex2025})
\end{itemize}

The rapid evolution of generative AI means capabilities once deemed impossible can shift dramatically in short timeframes---a phenomenon reminiscent of the pace of technological advancements in the 1990s. While historical parallels offer context, today’s advancements are uniquely driven by observations like Moore’s Law, which continue to underpin exponential growth in computing power and algorithmic efficiency~\cite{schaller1997moore}. For instance, we have witnessed the inference cost performing at GPT-3.5's level dropped 280 times within two years~\cite{aiindex2025}, GPT-4 significantly outperformed GPT-3.5 in financial literacy tests~\cite{niszczota2023gpt}, applications like ChatGPT and Gemini gained web-search functionality in late 2024, enabling them to have access to up-to-date information~\cite{openai2024chatgptsearch,google2024bardsearch}. OpenAI’s 2025 announcement of its artificial general intelligence (AGI) vision also underscores the accelerating and unpredictable trajectory of generative AI~\cite{altman2025agi}. To navigate through the generative AI realm, learners should update their knowledge of AI’s capabilities and limitations, as they are often temporary, not absolute.

\section{Discussion}

An AI literacy constructs review from across the past five years~\cite{almatrafi2024systematic} reveals that AI literacy guidelines are continually evolving--as are the learning interventions designed by educators and researchers. The unprecedented rise of generative AI since late 2022, and its exhibited characteristics, underscore the need to specify more nuanced learning goals towards generative AI as a subfield of AI literacy more broadly. A search for current university guidelines~\cite{harvard2023,ubc_genai2024,gwu_ai_guidelines,hku_ai2023}, and government documents~\cite{unesco_chatgpt2023,counseleu2023chatgpt} for generative AI policies did not fully cover the scope of needed broader learning objectives and considerations. As we systematically reviewed the literature, we also witnessed a disproportionate coverage of each guideline item in the literature. Specifically, guideline items G6: misinformation and falsified facts, G8: innate bias in outputs, and G11: broader social impact were covered in broader literature. The existing institutional policy documents' partial coverage and the disproportionate focus of academic literature underscore the contingency of a comprehensive and solid framework to define generative AI literacy to inform and guide future efforts. Notably, we deliberately excluded understanding the generative AI working mechanisms in our guidelines, as such technical information is unlikely to improve learners' ability to practically interact with generative AI. Due to the variance in target population, we encourage future researchers and practitioners to adapt our guidelines for specific contexts, such as for employees~\cite{cetindamar2022explicating}, middle school students~\cite{zhang2023integrating}, and families~\cite{druga20214as}. Different populations need different levels of understanding, and not all guideline items will be equally relevant to every group.

Unlike what is covered in the more established AI literacy competencies guidelines~\cite{long2020ai, ng2021conceptualizing, ng2021ai}, generative AI is a relatively new and rapidly evolving field~\cite{foster2023generative}, making it challenging to create an all-encompassing framework. In the meantime, we need to incorporate timely grey literature to complement the lengthy academic peer-review process, which allowed us to capture the rapidly evolving characteristics of generative AI. While there have been notable efforts in defining generative AI literacy~\cite{annapureddy2024generative}, chartering generative AI literacy assessment tests~\cite{jin2024glat}, there is still a need for more comprehensive and robust learning materials. We hope our guidelines will provide a more solid foundation for addressing the learning objectives. Future learning interventions could incorporate design considerations~\cite{long2020ai} in designing AI literacy interventions. Given the fast-paced evolution of generative AI, learning interventions need to be updated with the latest challenges that learners face to help learners interact with generative AI effectively, responsibly, and in an informed way.

\section{Conclusion and Future Work}

In this paper, we systematically analyzed and reviewed existing literature to charter 12 guideline items delineating the generative AI literacy guidelines, covering four key aspects. Our work aims to construct a holistic framework that serves as a starting point to scaffold future discourse across AI, learning science, and HCI communities and guide the development of learning interventions. We encourage educators, schools, companies, and organizations to adapt these guidelines into their own frameworks to meet the specific needs of their members---such as students and employees---enabling them to use generative AI in an ethical and informed manner. As generative AI becomes an inseparable part of daily life, significant work is needed to develop robust learning interventions that foster generative AI literacy. 

\bibliographystyle{ACM-Reference-Format}


\end{document}